\documentclass[aps,prl,twocolumn,nopacs]{revtex4}
\usepackage{natbib}
\usepackage[latin9]{inputenc}
\usepackage{amsmath}
\usepackage{amssymb}
\usepackage{graphicx}
\usepackage{esint}
\usepackage{braket}
\usepackage{multirow}
\usepackage{dcolumn}
\usepackage{color}
\usepackage[colorlinks,
            linkcolor=blue,
            citecolor=blue,
            anchorcolor=blue,
            urlcolor=blue,
            ]{hyperref}

\makeatletter

\newcommand{\unit}[1]{\ensuremath{\,\mathrm{#1}}}

\newcommand{\MHz}{\unit{MHz}}
\newcommand{\GHz}{\unit{GHz}}

\newcommand{\Yb}{\ensuremath{^{171}\mathrm{Yb}^+~}}
\newcommand{\avg}[1]{\ensuremath{\left\langle#1\right\rangle}}

\makeatother

\begin{document}
\title{Vacuum Measurements and Quantum State Reconstruction of Phonons}
\author{Dingshun Lv$^{1}$,Shuoming An$^{1}$, Mark Um$^{1}$, Junhua Zhang$^1$, Jing -Ning Zhang$^1$, M. S. Kim$^{2}$ \& Kihwan Kim$^{1}$}

\affiliation{$^{1}$Center for Quantum Information, Institute for Interdisciplinary Information Sciences, Tsinghua University, Beijing 100084, P. R. China  \\ $^{2}$QOLS, Blackett Laboratory, Imperial College London, SW7 2AZ, United Kingdom} 

\begin{abstract}
A quantum state is fully characterized by its density matrix or equivalently by its quasiprobabilities in phase space. A scheme to identify the quasiprobabilities of a quantum state is an important tool in the recent development of quantum technologies. Based on our highly efficient vacuum measurement scheme, we measure the quasiprobability $Q$-function of the vibrational motion for a \Yb ion {\it resonantly} interacting with its internal energy states. This interaction model is known as the Jaynes-Cummings model which is one of the fundamental models in quantum electrodynamics. We apply the capability of the vacuum measurement to study the Jaynes-Cummings dynamics, where the Gaussian peak of the initial coherent state is known to bifurcate and rotate around the origin of phase space. They merge at the so-called revival time at the other side of phase space. The measured $Q$-function agrees with the theoretical prediction. Moreover, we reconstruct the Wigner function by deconvoluting the $Q$-function and observe the quantum interference in the Wigner function at half of the revival time, where the vibrational state becomes nearly disentangled from the internal energy states and forms a superposition of two composite states. The scheme can be applied to other physical setups including cavity or circuit-QED and optomechanical systems. 
\end{abstract}

\maketitle
Reconstructing the state of a quantum system through measurements reveals all the statistical properties of the system. Thus schemes to reconstruct a quantum state are useful, for example, to ensure the quantum state generated and to test the fidelity of quantum operations. The quantum state is equivalently represented by its density matrices or quasiprobability functions in phase space \cite{Glauber69}. Among the quasiprobability functions, the Wigner function has been used mainly for the study of non-classicality of the state, which is manifested by negativities \cite{Kenfack04}. The $Q$-function has been used to study the essence of the dynamics of a quantum-state evolution in phase space \cite{Phoenix88,Banacloche90,Banacloche91,Eiselt89,Eiselt91,Auffeves03,Raimond05,7Haroche,4BEC,Schoelkopf13a}. Recently, there have been many developments in reconstructing the state of a quantum field in various physical systems including photonic systems \cite{Faridani93,Raymer09}, atomic systems \cite{Kurtsiefer97}, molecular systems \cite{Walmsley95}, trapped ion systems \cite{Monroe96,Leibfried96,Home15,Dzmitry15,Urabe15}, cavity-QED \cite{Deleglise08} and circuit-QED systems \cite{Cleland09,Schoelkopf13b}, which are mostly related to the reconstruction of Wigner functions by the parity measurements of the states.   

The $Q$-function requires only the measurement of the vacuum component of the field, which looks relatively easy to implement. The reconstruction of the $Q$-function also does not require a heavy numerical process as the probability of the state being in the vacuum is the value of the $Q$-function in each point of phase space. However, the measurement of the vacuum state is not straightforward. In a cavity, as an example, the existence of a photon can be detected by an atomic state through an atom-photon interaction like Jaynes-Cummings model (JCM)\cite{Jaynes63}. If the cavity has no photons, an atom initially prepared in its ground state will remain there forever. However, by merely measuring the atom in its ground state we cannot say that the cavity is empty because the Rabi oscillations periodically bring the atom back to the ground state even with many photons in the cavity. The oscillation frequency depends on how many photons are present in the cavity. The authors in Refs. \cite{Auffeves03,Raimond05,7Haroche} demonstrated a scheme of vacuum measurement that works for their particular cavity-QED system. For the circuit-QED system \cite{Schoelkopf13a}, a measurement of the vacuum component and the $Q$-function reconstruction was demonstrated based on the system-specific strong-nonlinear coupling between the cavity mode and the artificial atom. Recently a generic scheme of the vacuum measurement was proposed for the cavity-QED system with the standard JCM coupling \cite{Jeffers13}. 

While it is desirable to find the $Q$-function and the Wigner function based on one set of measurement, this has not been achieved due to the measurement inefficiencies of the vacuum state. Here, we report a generic and efficient detection of vacuum with 98.5$(\pm 0.3)\%$ efficiency for the phononic states in the vibrational mode of a harmonic trap, which is realized by the adiabatic passage schemes \cite{Piet16,Shuoming15,UmMark15} based on counter-diabatic methods \cite{Rice03,Berry09,24LaudauZener}. The demonstrated adiabatic passages have been significantly improved in order to measure the vacuum component of a reasonably large phonon state up to $\avg{n}\sim 25$ phonons. Typically, in trapped ion systems, phonon number distributions are measured by the Fourier transformation of the phonon-number dependent Rabi oscillations \cite{Meekhof96,Leibfried03,Haffner08}. For this, a long observation time is necessary. Our scheme does not require such a long observation time, but at each measurement, we obtain a binary result; the vacuum or the complementary states. We simply need to repeat the measurement sequence for the probabilities of the vacuum state, which by nature of the measurement excludes the negativity problem of the $Q$-function shown in Ref. \cite{Schoelkopf13a}. We efficiently measure the vacuum probabilities in phase space to observe the dynamics of the JCM field. The measured $Q$-function is highly accurate, which enables us to reconstruct the density matrix and the Wigner function by its deconvolution. We show a good agreement of our quasiprobabilities with the theoretical predictions.

\begin{figure}[hp]
\includegraphics[width=0.45\textwidth]{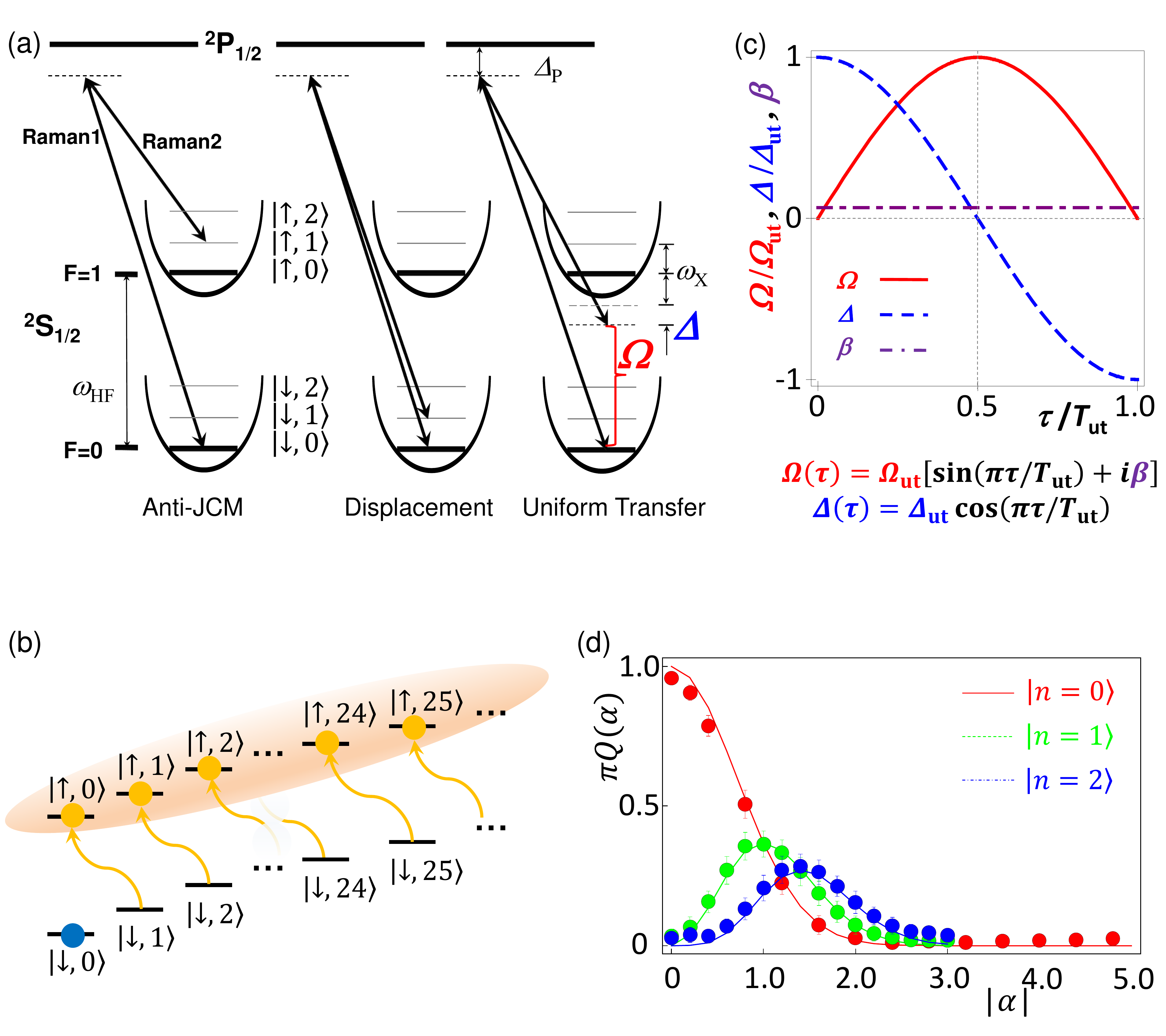}\\
\caption{{\bf Raman laser schemes and the vacuum measurement.} (a) The Hilbert space of the system is comprised of the direct product of qubit states $\left\{ \ket{\downarrow}, \ket{\uparrow} \right\}$ and phonon number states $\left\{ \ket{n=0}, \ket{1}, \ket{2},... \right\}$. Raman laser beams, which have $\sigma_{-}$ polarization and are detunned by $\Delta_{\rm p} \approx$ 12.9 THz from the $P_{1/2}$ manifold, perform anti-JCM, displacement operation and the vacuum measurement by adjusting their beating frequencies (see Method A). (b) The vacuum component is measured by transferring the population of $\ket{\downarrow, n}$ to that of $\ket{\uparrow, n-1}$ for any value of $n$ at the same duration of pulse. The atom remaining in no fluorescence state $\ket{\downarrow}$ indicates the phononic state being in $\ket{0}$. (c) The uniform transfer for any phononic state $\ket{n}$ to $\ket{n-1}$ is accomplished by the scheme of shortcuts to the adiabaticity, where $\Omega_{\rm ut} = (2 \pi) 22.7$ kHz, $\beta = 0.075$, $\Delta_{\rm ut} = 1.9 \Omega_{\rm ut}$ and the total duration $T_{\rm ut} = 198.2 \mu$s. (d) $Q$-function of the phononic Fock state $n=0,1,2$ depending on the amount of displacement $|\alpha|$. The points with error bars are the experimental results while the dashed lines are by the theory. The error bars are obtained by the standard deviation of the quantum projection noise with 100 repetitions. \label{fig:Qmeasurement}}
\end{figure}

{\bf Jaynes-Cummings dynamics}\\
The JCM is one of the most fundamental interaction models in quantum mechanics \cite{Jaynes63}, where a single two-level atom resonantly interacts with a single-mode field. The JCM has enabled theoretical and experimental investigations of the basic properties of quantum electrodynamics such as Rabi oscillations of the energy transfer between the two subsystems and collapses and revivals of the oscillations \cite{Eberly80}. More recently, the model has been widely studied for its rich properties of quantum control, coherent superposition and entanglement which are closely related to the current development of quantum technology. In order to see the nonclassical effects due to quantum interaction, the JCM is often studied with the state initially prepared in a coherent field and the atom in its energy eigenstate. It has been shown that the field and the atom are entangled \cite{Phoenix88} as soon as the interaction starts, but at a certain time they are nearly disentangled to bring the field into a superposition of two coherent states of a $\pi$ phase difference \cite{Banacloche90,Banacloche91}. Earlier, Eiselt and Risken \cite{Eiselt89,Eiselt91} showed that the Gaussian probability distribution of the initial coherent state in phase space breathe at the initial points of interaction, reflecting the Rabi oscillations. Then the Gaussian peak bifurcates to travel around a circle in opposite direction in phase space. The bifurcation is a consequence of quantum nature of interaction and was experimentally probed through the measurement of field phase distribution \cite{Auffeves03,Raimond05}. However, the full reconstruction of the dynamics of the JCM field has not been experimentally demonstrated.  

{\bf Experimental setup}\\
We employ the vibrational mode of a single trapped ion \Yb in a harmonic potential with the frequency of $\omega_{\rm X}$ = (2$\pi$) 2.8 \MHz. We encode the qubit state into two hyperfine states $\ket{F=1,m_{F}=0}\equiv\ket{\uparrow}$ and $\ket{F=0,m_{F}=0}\equiv\ket{\downarrow}$ of the $S_{1/2}$ manifold with the transition frequency $\omega_{\rm HF} = (2 \pi) ~ 12.6428$ \GHz. We realize the JCM or anti-JCM by applying a pair of counter-propagating Raman beams that have the frequency differences of ($\omega_{\rm R1}-\omega_{\rm R2})=\omega_{\rm HF}\mp\omega_{\rm X}$, respectively, as shown in Fig. \ref{fig:Qmeasurement}(a). In the interaction picture, the Raman laser interactions can be described by the following JCM and anti-JCM Hamiltonians
\begin{eqnarray}
\hat{H}_{\rm JC} \left(\phi \right)&=& \frac{\hbar\eta\Omega}{2}\left(\hat{a}\hat{\sigma}_{+} \mbox{e}^{i \phi} +\hat{a}^{\dag}\hat{\sigma}_{-}\mbox{e}^{-i \phi}\right), \nonumber \\ \hat{H}_{\rm aJC} \left(\phi \right) &=& \frac{\hbar\eta\Omega}{2}\left(\hat{a}^{\dag}\hat{\sigma}_{+} \mbox{e}^{i \phi}+\hat{a}\hat{\sigma}_{-} \mbox{e}^{-i \phi}\right). 
\label{JCnAJC}
\end{eqnarray}
Here, $\hat{a}^{\dag}$ and $\hat{a}$ are the phonon creation and annihilation operators, $\hat{\sigma}_{+} \left(\hat{\sigma}_{-}\right)=\ket{\uparrow}\bra{\downarrow} (\ket{\downarrow}\bra{\uparrow})$ the spin-raising (lowering) operator, $\Omega$ the vacuum Rabi frequency of (anti-)JCM and $\eta=\Delta k \sqrt{\hbar/ M\omega_{\rm X}}$ the Lamb-Dicke parameter with $\Delta k$ the net wave-vector of the Raman laser beams, $M$ the mass of the \Yb ion and $\phi$ the phase difference of the Raman laser beams. The JCM and anti-JCM dynamics share more or less the same behavior. For technical reasons, we perform experiments for the anti-JCM interaction. 

\begin{figure*}[ht]
\includegraphics[width=1.0\linewidth]{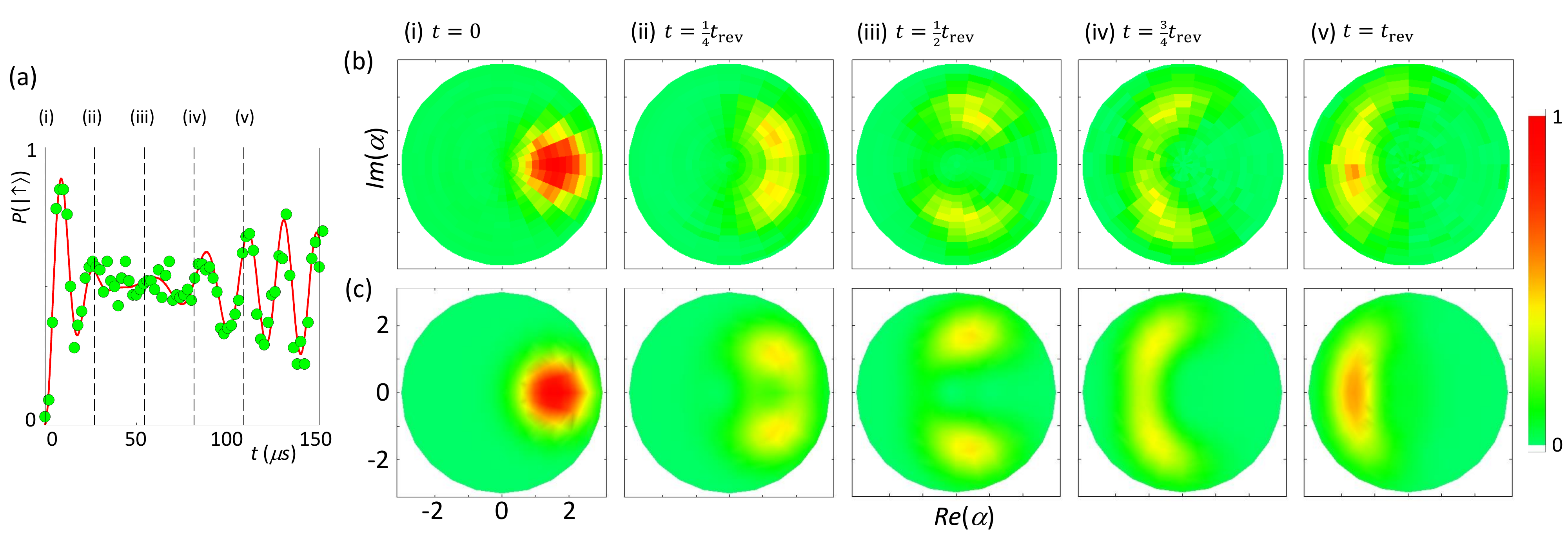}\\
\caption{{\bf The time evolution of the $Q$-function for an initial coherent state under anti-JCM interaction.} (a) Collapse and revival of the Rabi oscillation signal, (b) experimentally measured and (c) numerically calculated $Q$-functions of the phonon field with the initial coherent state $\ket{\beta=1.62(0.05)}$ depending on the duration of anti-JCM interaction. (a) $P(\ket{\uparrow})$ is the probability of detecting the atom in $\ket{\uparrow}$. The points are obtained after 100 repetitions. The solid line is from fitting the data with $\sum_{n=0} \frac{1}{2}\left[1-\mbox{e}^{-\gamma t}\cos \left( \sqrt{n+1} \eta\Omega t \right)\right]$, where $\gamma$ is the empirical decay constant. At (b) and (c), the time for each snapshot of the $Q$-functions are labeled as (i)-(v) in the unit of the revival time $t_{\rm rev}$, where $t_{\rm rev}=\frac{4\pi |\alpha|}{\eta\Omega}= 108.8~\mu$s. In (b), each $Q$-function is obtained from 100 repetitions of the vacuum measurements after 384 different displacements, where the amplitude and the phase of displacement $|\alpha| \mbox{e}^{i \varphi}$ are scanned from 0 to 3.0 with the step size of 0.2 and from 0 to $2\pi$ with the 24 steps, respectively (see also Supplementary Movie).\label{fig:QF}}
\end{figure*} 

{\bf Efficient vacuum measurement}\\
The essence of the vacuum-component measurement is in the realization of the uniform population transfer of $\ket{\downarrow, n} \rightarrow \ket{\uparrow, n-1}$ for any $n$ as shown in Fig. \ref{fig:Qmeasurement}(b). After the uniform transfer, all the phonon states except the vacuum component $\ket{n=0}$ are in the bright electronic state $\ket{\uparrow}$, which emits photons during the standard fluorescence detection sequence. Therefore, the atom being in the dark electronic state $\ket{\downarrow}$ after the uniform transfer indicates the phonon state in the vacuum. By measuring the vacuum probability of the state after displacing it by $\alpha$, we can directly measure the $Q$-function $Q(\alpha)$. The $Q$-function allows to study the core of the dynamics of a quantum state in phase space and has well been a preferred choice of study theoretically \cite{Phoenix88,Banacloche90,Banacloche91,Eiselt89,Eiselt91} and experimentally \cite{7Haroche,Auffeves03,Raimond05,4BEC,Schoelkopf13a}. The definition of the $Q$-function is $Q(\alpha)=\frac{1}{\pi} \bra{0}\hat{D}^{\dagger}(\alpha)\hat{\rho}\hat{D}(\alpha)\ket{0}$, where $\hat{D}(\alpha)$ is the displacement operator \cite{Barnett97}; the value of the $Q$-function is merely the weight of the vacuum component of a given state once it is displaced in phase space. 

In general, the frequency of the Rabi oscillations between $\ket{\downarrow, n}$ and $\ket{\uparrow, n-1}$ has $\sqrt{n}$ dependency due to the nature of JCM coupling. To accomplish the uniform transfer, we basically apply an adiabatic passage, but in much shorter time than what is required for the adiabatic evolution; the so-called shortcuts to adiabaticity \cite{Rice03,Berry09,24LaudauZener,Shuoming15,UmMark15}. Here, as shown in Fig. \ref{fig:Qmeasurement}(c), the detuning $\Delta \equiv (\omega_{\rm R1}-\omega_{\rm R2})-(\omega_{\rm HF}-\omega_{\rm X} )$ and the amplitude $\Omega$ of Raman laser beams are swept by $\Delta(t)=\Delta_{\rm ut} \cos(\pi t/T_{\rm ut})$ and $\Omega (t) = \Omega_{\rm ut} \left[\sin(\pi t/T_{\rm ut}) + i\beta \right]$, where $i \beta$ is the counter-diabatic field that is applied at a constant amplitude with a 90 degree out of phase with the driving field to suppress excitations during the fast evolution \cite{Rice03,Berry09,24LaudauZener,Shuoming15,UmMark15}. We evaluate the reliability of the uniform transfer by performing the $Q$-function measurements for the phonon number states $\ket{n=0,1,2}$ as shown in Fig. \ref{fig:Qmeasurement}(d). Here, we prepare the phonon number states $\ket{n=0,1,2}$ and displace them along one direction in phase space. We note that we do not observe serious imperfection over the quantum projection noise. 

{\bf $Q$-function reconstruction for JCM dynamics}\\
It was found that the atom and the field in the JCM or anti-JCM are nearly disentangled during the course of interaction if the atom is initially prepared in a superposition of $\ket{\uparrow}$ and $\ket{\downarrow}$ and the field is initially in the coherent state $\ket{\alpha}$ of its amplitude $\alpha$ with $|\alpha|\gg 1$. Let us consider the initial state of the atom $\ket{\Psi_A^{\pm}}=\left(\ket{\uparrow}\mp i\ket{\downarrow}\right)/\sqrt{2}$. By the interaction (\ref{JCnAJC}), the atom-field state evolves to  $|\Psi_{A-P}^{\pm}(t)\rangle=|\Psi_P^{\pm}(t)\rangle\otimes|\Psi_A^{\pm}(t)\rangle$ \cite{Buzek92}, where
\begin{eqnarray}
\ket{\Psi_A^{\pm}(t)} &=& \left(\mbox{e}^{\pm i \pi t/t_{\rm rev}}\ket{\downarrow}\mp i\ket{\uparrow}\right)/\sqrt{2} \label{psiA}\\
\ket{\Psi_P^{\pm}(t)} &=& \exp\left(\mp i t \frac{\eta \Omega \sqrt{\hat{n}}}{2}\right)\ket{\alpha \mbox{e}^{\pm i \pi t/t_{\rm rev}}}.
\label{psiP}
\end{eqnarray}
From this, it is clear that if the atom is prepared in its ground state $\ket{\downarrow}$ $\left[=\left(\ket{\Psi_A^{-}}-\ket{\Psi_A^{+}}\right)/\sqrt{2}i \right]$, the atom-phonon state will be in the superposition of $\ket{\Psi_{A-P}^{\pm}(t)}$. The phonon state will rotate in phase space, where $t_{\rm rev}=\frac{4\pi |\alpha|}{\eta\Omega}$ is the corresponding revival time.

In the experiment, we prepare a coherent state of $\beta=1.62(0.05)$ with the internal state $\ket{\downarrow}$ by displacing the $\ket{n=0}$ state after the standard Raman-sideband ground-state cooling (see Method A). Then we apply Raman laser beams for the anti-JCM interaction and observe the dynamics of the atom and the field. For the internal state of the atom, we measure the probability of being in the $\ket{\uparrow}$ state, $P\left({\ket{\uparrow}}\right)$ by the standard fluorescence detection scheme. For the field, we choose five different interaction times $t=(0,\frac{1}{4}, \frac{1}{2}, \frac{3}{4}, 1)~t_{\rm rev}$ in the anti-JCM. After the interaction time $t$, we displace the state by $\alpha$ and trace over the internal degree of freedom by the standard optical pumping sequence, which does not produce any noticeable change of the phonon distribution (see Method B and Supplementary Fig. 1). Then we measure the vacuum component to reconstruct $Q(\alpha)$. 

Figs. \ref{fig:QF}(b) and (c) show the experimental and theoretical time evolution of the initial coherent state under the anti-JCM interaction. The theoretical results are obtained by the numerical simulation of the master equation of anti-JCM Hamiltonian including experimental imperfections \cite{Shuoming15}. At $t=0$, $P({\ket{\uparrow}})=0$ and $Q(\alpha)$ is Gaussian, which represents the coherent state. At time $t= t_{\rm rev}/4$, while the Rabi oscillations begin to collapse, the Gaussian peak splits into two, which can be understood by the separation of two atom-phonon states $\ket{\Psi_{A-P}^{\pm}}$. The two components of the atom-phonon entangled state evolve in the opposite phases as shown in Eqs. (\ref{psiA}) and (\ref{psiP}). At the half revival time $t= t_{\rm rev}/2$, the two atomic states in Eq. (\ref{psiA}) become identical except the global phase, which results in disentanglement of the atomic state from the phonon state (see also Fig. \ref{fig:WF}). In the $Q$-function, the phonon state shows two clearly separated peaks that are located at the opposite phases in phase space. This can be understood as the superposition of two coherent states \cite{Buzek92}. Further evolution of the phonon state is shown in Figs. \ref{fig:QF}(iv) and (v). At the revival time $t= t_{\rm rev}$, the two phonon peaks merge at the opposite position of the initial coherent state, which causes the revival of the Rabi oscillations (see also Supplementary Fig. 2 and Discussion). Due to the quadratic phase term in Eq. (\ref{psiP}), the amplitude of the Rabi oscillations is reduced. 

\begin{figure}
\includegraphics[width=1.0\linewidth]{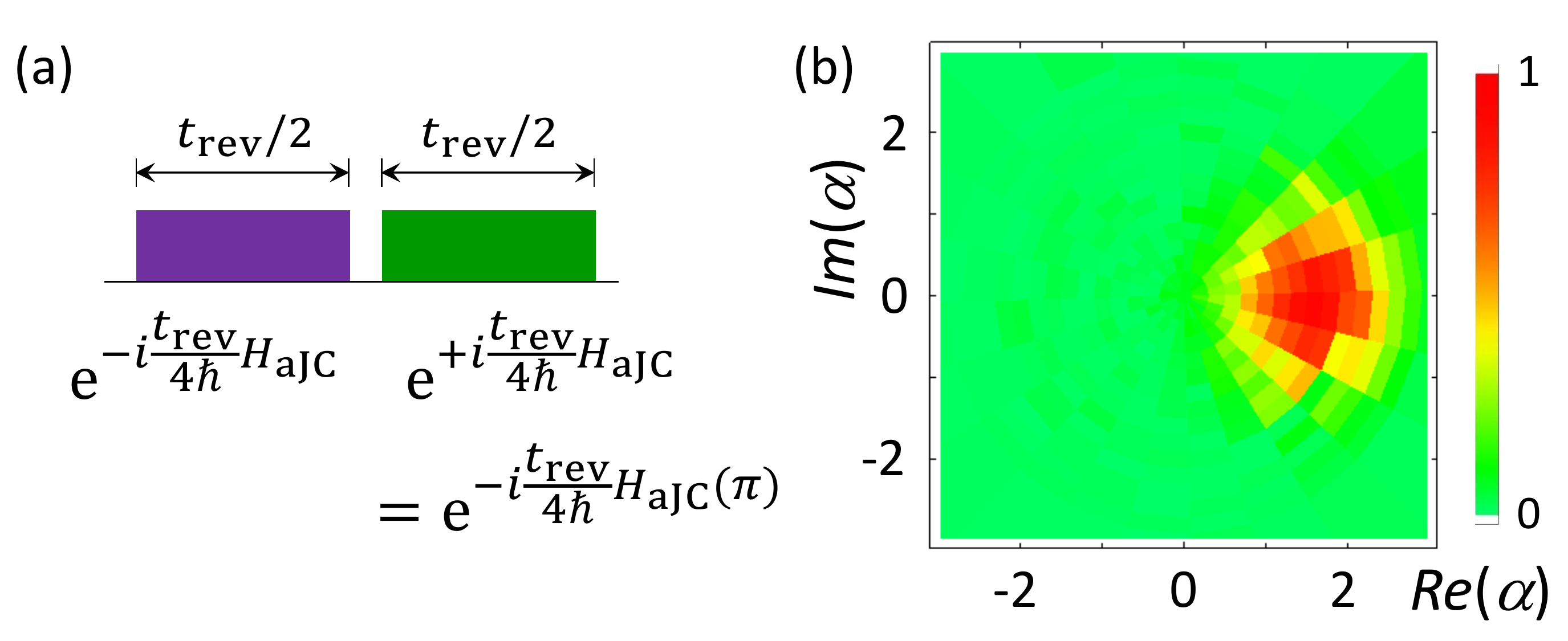}\\
\caption{(a) Generalized echo-sequence time reversal of anti-JCM evolution for the interaction time $t=t_{\rm rev}/2$. The $\phi=\pi$ phase of the anti-JCM Hamiltonian produces the negative sign; $H_{\rm aJC}(\pi)=-H_{\rm aJC}(0)$, which performs the time reversal operation. (b) The measured $Q$-function of the phononic state after time reversal operation of (a). The total number of measurements for the $Q$-function reconstruction is same as that in Fig. \ref{fig:QF}. \label{fig:QF-reverse}}
\end{figure}

{\bf Time-reversal operation}\\
In order to confirm the whole dynamics keeping coherence, we perform the time reversal operation, which forces the phonon state evolved under the anti-JCM interaction to retrace its past trajectory in the opposite direction by the generalized echo scheme \cite{echo02}. For the echo method, we introduce a $\pi$ phase shift in the second half of the anti-JCM interaction, $i.e.$, $\mbox{e}^{-i \frac{t}{2 \hbar} H_{\rm aJC}(\pi)}= \mbox{e}^{+i \frac{t}{2 \hbar} H_{\rm aJC}}$. The process is called time-reversal as in Ref. \cite{14TimeReversal}. We apply the reverse process at the half revival time $t=t_{\rm rev}/2$ and observe that the state is brought back to the initial coherent state at the time $t=t_{\rm rev}$ with the fidelity of 0.914(4) through the $Q$-function measurement shown in Fig. \ref{fig:QF-reverse} (see also Method C). Since keeping the coherence of the interaction is at the heart of the time reversal, our result of time reversal clearly confirms that the process occurs in quantum regime. We also study another way of reversing the anti-JCM by applying the JCM (see Supplementary Fig. 3 and Discussion). \cite{15combineAJCandJC}.  

\begin{figure}
\includegraphics[width=1.0\linewidth]{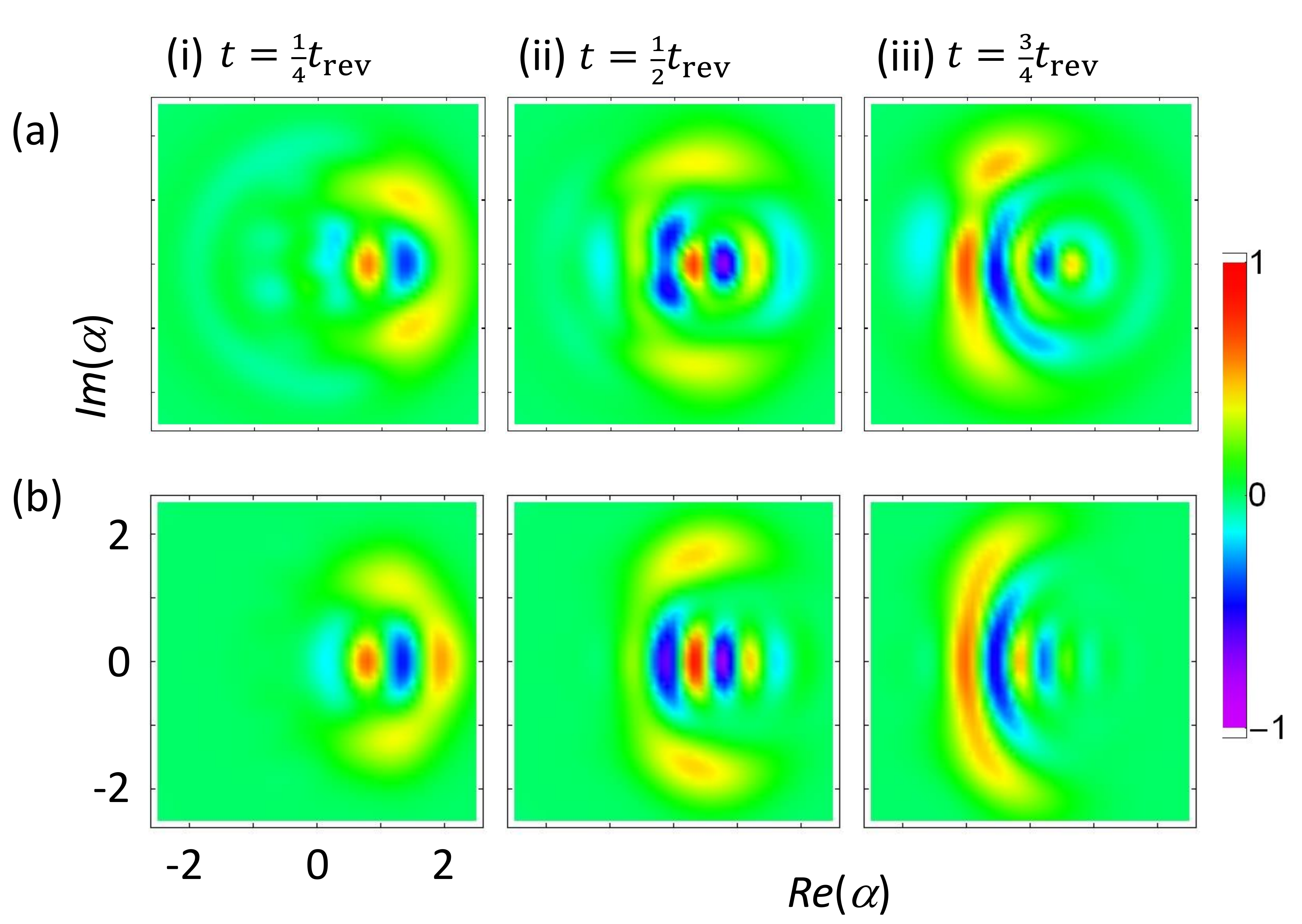}\\
\caption{ The Wigner function reconstruction from the $Q$-function at various times of the anti-JCM evolution, $t=\frac{1}{4}, \frac{1}{2}, \frac{3}{4} t_{\rm rev}$. The negativities of the Wigner functions indicate the emergence of nonclassical state during the dynamic evolution. (a) The Wigner functions are reconstructed from the density matrix obtained by deconvoluting the experimentally measured values of the $Q$-functions shown in Fig. \ref{fig:QF}(b). The density matrices are reconstructed by deconvoluting the $Q$-functions with the convex-optimization (see Method C). We note that we use the data $|\alpha|\leq 1$ for the optimum fidelity and it is necessary to use proper initial guess of the density matrix for the convergence of the deconvolution. (b) The Wigner functions are directly obtained by the numerical calculation of the anti-JCM dynamics.}\label{fig:WF}
\end{figure}

{\bf Wigner function reconstruction from $Q$-function}\\
In addition to the time-reversal process, we demonstrate the coherence property by detecting non-classicality generated during the evolution, in particular, interferences of the composite states of the two peaks in phase space. For this purpose, we reconstruct the Wigner function from our measured $Q$-function. We first find the density matrix by deconvoluting the $Q$-function by the convex-optimization (see Method C) and reconstruct the Wigner function from the density matrix. Fig. \ref{fig:WF}(a), which is reconstructed from the experimental data of the $Q$-function measurement, clearly manifests interference patterns of the composite states at the half revival time and negativities in other interaction times. The experimental reconstruction of the Wigner function of Fig. \ref{fig:WF}(a) is in good agreement with the direct theoretical reconstruction of the Wigner function shown in Fig. \ref{fig:WF}(b). We also obtain the purities $\mbox{Tr} \left(\rho^2 \right)$ of the states from the experimentally reconstructed density matrix. At $t=t_{\rm rev}/2$, the purity is $0.82(0.05)$, which indicates the phonon state is not entirely pure, possibly because of its entanglement with the internal state (see Supplementary Fig. 2). Theoretical studies \cite{Banacloche90} suggest that the purity reaches ideally at unity as the size of initial coherent state increases.

We have shown a highly efficient scheme to detect the vacuum which is used to reconstruct the dynamics of the JCM field state. The efficient measurement of the $Q$-function enables us to reconstruct the Wigner function. To our knowledge, this is the first demonstration of the Wigner function reconstruction from the vacuum measurement. Our developed technique of the $Q$-function measurement can be used to probe other dynamics of the phonon field including Kerr dynamics. In our experimental demonstration, the size of the initial coherent state $\ket{\beta}$ could be increased by improving the ion trap system. The main limitation of the current demonstration comes from the unreliable displacement operation above $\alpha \approx 4.8$, which is caused by heating of the phonon mode and going outside Lamb-Dicke regime of the phonon state. The reduction of an order of magnitude in the heating rate would allow us to reach an order of magnitude large phonon number state. Our approach is generic and would also be applied to other physical platforms that have a Jaynes-Cumming type of coupling including opto-mechanics and circuit-QED system.

\section*{Acknowledgements}
This work was supported by the National Basic Research Program of China under Grants No. 2011CBA00300 (No. 2011CBA00301), the National Natural Science Foundation of China 11374178, 11574002 and 11504197. M.S. Kim thanks Matteo Paris, Girish Agarwal and Werner Vogel for the discussions on the Wigner function reconstruction and was supported by the UK EPSRC (EP/K034480/1) and the Royal Society.


\subsection*{Additional information}
The authors declare no competing financial interests. Reprints and permissions information is available online at www.nature.com/reprints. Correspondences should be addressed to M.S.Kim (m.kim@imperial.ac.uk) and K.Kim (kimkihwan@mail.tsinghua.edu.cn).






\end{document}